\documentclass[12pt,preprint]{aastex}

\shortauthors{Bailin}
\shorttitle{Sgr and the Milky Way warp}

\begin{document}

\title{Evidence for coupling between the Sagittarius dwarf galaxy
and the Milky Way warp}

\author{Jeremy Bailin}
\affil{Steward Observatory, University of Arizona}
\affil{933 North Cherry Ave, Tucson, AZ 85721, USA}
\email{jbailin@as.arizona.edu}

\begin{abstract}%
Using recent determinations of the mass and orbit of Sagittarius,
I calculate its orbital angular momentum.
From the latest observational data, I also calculate the
angular momentum of the Milky Way's warp.
I find that both angular
momenta are directed toward $l\approx 270\degr$, $b=0\degr$,
and have magnitude 2--$8\times 10^{12}~M_{\Sun}~\mathrm{kpc~km~s^{-1}}$,
where the range in both cases reflects uncertainty in the mass.
The coincidence of the angular momenta is suggestive of a
coupling between these systems.
Direct gravitational torque of Sgr on the disk is ruled out
as the coupling mechanism.
Gravitational torque due to a wake in the halo and the impulsive deposition
of momentum by a passage of Sgr through the disk
are still both viable mechanisms pending better simulations
to test their predictions on the observed Sgr-MW system.%
\end{abstract}


\section{Introduction}
\label{intro}
\begin{figure}
\plotone{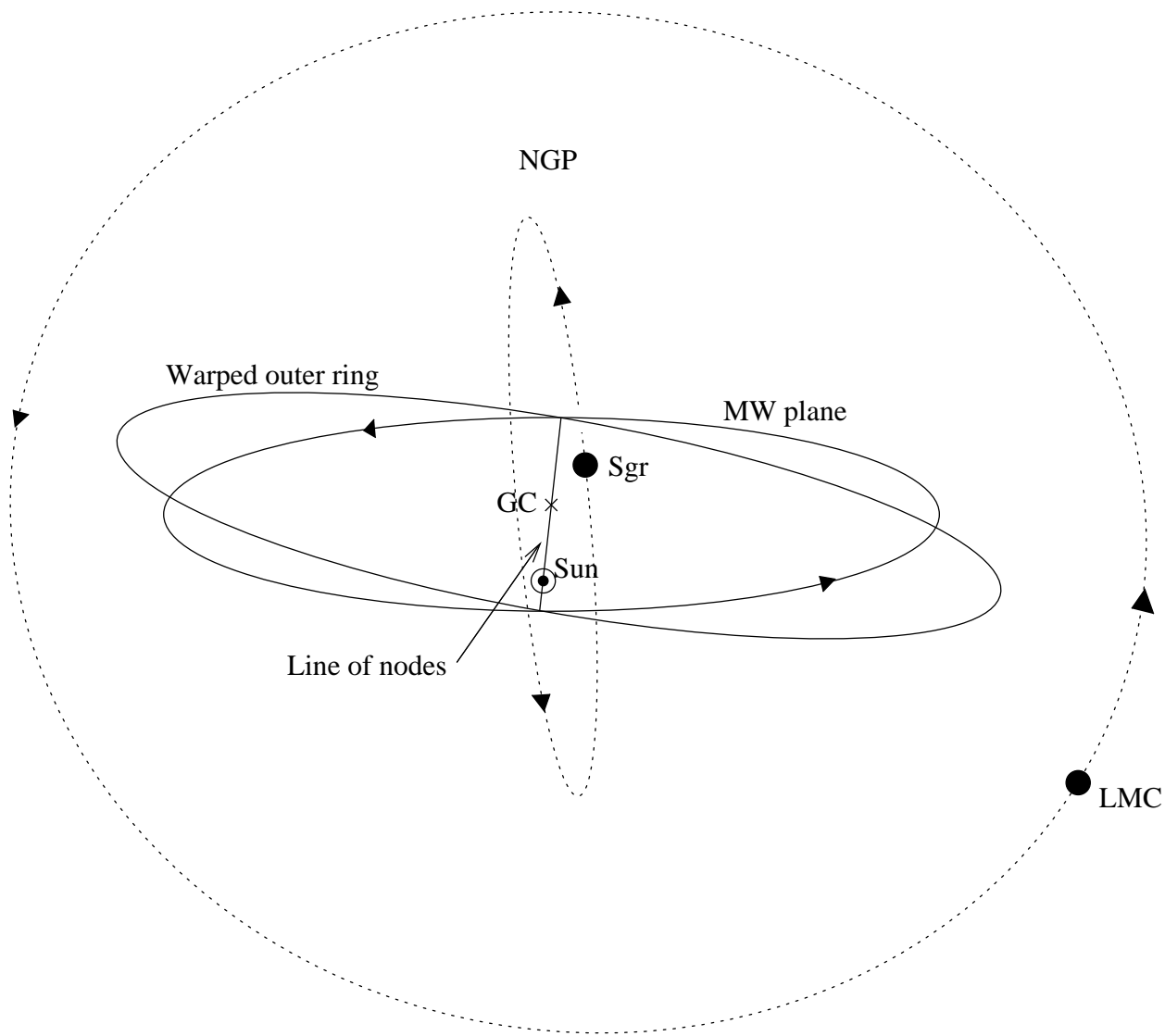}
\caption{Schematic drawing of the plane of the Milky Way, the
Galactic center (GC), the line
of nodes of the warp, and the orbits of Sagittarius (Sgr) and the Large
Magellanic Cloud (LMC). The plane of Sagittarius's orbit intersects
the line of nodes and is orthogonal to the plane of the LMC's
orbit. Not to scale.\label{schematic}}
\end{figure}

The disk of the Milky Way is warped like an integral sign, rising
above the plane on one side and falling below the plane on the other.
This warp is seen both in maps of neutral hydrogen
\citep[e.g.,][]{diplas and savage91} and in the stellar distribution
(Reed 1996; Drimmel, Smart, \& Lattanzi 2000; L\'opez-Corredoira
et al. 2002b).
The Sun lies along the line of nodes of the warp, where tilted
outer rings cross the inner plane (see Figure~\ref{schematic}).

Despite their tendency to disperse when isolated
\citep{hunter and toomre69},
warps are common in external galaxies
(Bosma 1981; Briggs 1990; Christodoulou, Tohline, \& Steiman-Cameron 1993;
Reshetnikov \& Combes 1998).
This has driven many authors to search for universal mechanisms to excite
or maintain warps \citep[see][for a review]{binney92}. Many of these
proposed mechanisms rely on the dark halo to
either stabilize warps as discrete bending modes within the halo
(Sparke \& Casertano 1988; but see also Binney, Jiang, \& Dutta 1998),
or to provide the torque necessary to create the warp 
\citep{ostriker and binney89,debattista and sellwood99,ideta et al00}.
Other proposed mechanisms
include the infall of intergalactic gas
(L\'opez-Corredoira, Betancort-Rijo, \& Beckman 2002a),
magnetic fields
(Battaner, Florido, \& Sanchez-Saavedra 1990), 
and interactions with
satellite galaxies \citep[e.g.,][]{huang and carlberg97}.

Each of these mechanisms can, in particular circumstances, produce
realistic-looking galactic warps. Although no single mechanism appears
universal enough to account for all warps, the evolution toward
a bending mode (even when no discrete mode exists) appears
enough like an observed warp \citep{hofner and sparke94} that
warping may be a generic response of disks to the individual
perturbations they experience. In this case, we should look
at individual warped galaxies for specific evidence of particular
perturbations that explain their warps rather than search for
a universal mechanism that may not exist.

The Magellanic Clouds have been proposed as the perturbation
responsible for the Milky Way's warp.
While \citet{hunter and toomre69} found that the tidal distortion from
the clouds alone is not sufficient to cause the observed warp,
\citet{weinberg98} proposed that orbiting satellites could set
up wakes in the Milky Way's halo which could provide the necessary
torque. \citet{tsuchiya02} performed self-consistent simulations
of such a system and confirmed that for a sufficiently massive
halo ($2.1\times10^{12}~M_{\Sun}$), the magnitude of the torque
can be increased enough to cause a warp of the same magnitude
as the Milky Way's.

The Magellanic Clouds orbit about the center of
the Galaxy in a direction orthogonal to the line of nodes,
i.e., near the line of maximum warp (see Figure~\ref{schematic}).
Garc\'ia-Ruiz, Kuijken, \& Dubinski (2000)
demonstrated 
that the warp caused by a satellite
will have its line of nodes oriented \textit{along} the satellite's orbit.
A simple way of understanding this result is to recognize that
a torque is a transfer of angular momentum, and therefore the
disk will acquire angular momentum along the same axis as the
orbital angular momentum of the satellite which is providing
the torque, and tilt toward that axis.
Therefore, the Magellanic Clouds are a bad candidate for
producing the Milky Way warp.

The orbital plane of the Sagittarius dwarf galaxy
(Ibata, Gilmore, \& Irwin 1994)
does intersect the line of nodes, suggesting that it may be
a good candidate for producing the Milky Way warp \citep{lin96}.
It is located behind the Galactic bulge and is on a nearly polar
orbit \citep{ibata et al97}.
\citet{ibata and razoumov98} performed simulations which suggest
that the passage of a sufficiently massive Sgr 
($5\times 10^9~M_{\Sun}$) through the disk could produce a warp.
Alternatively, its gravitational tides or the tides of a wake
it produces
in the dark halo could exert a warp-inducing torque on the disk.

If Sgr is responsible for the warp, its angular momentum will
be coupled to that of the warp. In this Letter, I calculate
the orbital angular momentum of Sgr, along with the component of the
Milky Way disk's angular momentum which does not lie in the
common plane. I show that they have the same
direction and the same magnitude. As there is
no \textit{a priori} reason to expect them to be within orders of magnitude
of each other, this is evidence that Sgr is coupled to the
warp, and therefore responsible for it.

\section{The angular momentum of satellite galaxies}
\label{Lsgr section}
\begin{deluxetable}{lcc}
\tablecaption{Properties of the Sagittarius dwarf\label{sgr properties}\label{Lsgr table}}
\startdata
\sidehead{\citet{ibata et al97}:}\\
Galactic coordinates	& \multicolumn{2}{c}{$l=5.6\degr$, $b=-14\degr$}\\
Galactocentric distance	& \multicolumn{2}{c}{$16\pm2$~kpc}\\
Space motion $(U,V,W)$	& \multicolumn{2}{c}{$(232,0,194)\pm60~\mathrm{km~s^{-1}}$}\\
Galactocentric radial velocity	& \multicolumn{2}{c}{$150\pm60~\mathrm{km~s^{-1}}$}\\
Galactocentric tangential velocity	& \multicolumn{2}{c}{$270\pm100~\mathrm{km~s^{-1}}$}\\
\sidehead{Derived angular momentum:}\\
Assumed mass ($10^9~M_{\Sun}$)	& 0.4	& 2.0 \\
Angular momentum ($10^{12}~M_{\Sun}~\mathrm{kpc~km~s^{-1}}$)	& $1.7\pm0.6$	& $8.6\pm3.4$ \\
Direction	& \multicolumn{2}{c}{$l=276\degr$, $b=0\degr$} \\
\enddata
\end{deluxetable}
The position, distance, and motion of Sagittarius are given
in Table~\ref{sgr properties}, along with estimates of its
mass and orbital angular momentum. The angular momentum can
range between 1.7~and $8.6\times 10^{12}~M_{\Sun}~\mathrm{kpc~km~s^{-1}}$
and is directed toward $l=276\degr$, $b=0\degr$.

The major uncertainty in this calculation is
the determination of the mass.
\citet{ibata and lewis98}
argued that in order for the satellite to have survived
to the present day, it
must have a massive extended dark matter halo and a total
$M/L\sim 100$ in solar units \citep{ibata et al97}.
However, \citet{helmi and white01} found viable models
with more moderate masses ranging from $4.66\times 10^8~M_{\Sun}$
for a purely stellar model to $1.7\times 10^9~M_{\Sun}$ for
their model with an extended dark matter envelope
\citep[see also][who find that if the original mass of
Sgr was large enough for dynamical friction to be important,
the majority of the mass would have been stripped off after
a Hubble time leaving a current mass of $1$--$3\times 10^9~M_{\Sun}$]{jiang and binney00}.
The properties of \citet{helmi and white01}'s models seem
most in agreement with the expected properties of dwarf
spheroidal galaxies, and therefore I adopt $0.4$--$2.0\times 10^9~M_{\Sun}$
as the range of possible masses of the Sagittarius dwarf.

\begin{deluxetable}{lccl}
\tablecaption{Angular momenta of the Milky Way warp and some
Milky Way satellites\label{satellites table}}
\tablehead{ & \colhead{Angular momentum $(M_{\Sun}~\mathrm{kpc~km~s^{-1}})$} &
\colhead{Direction} & \colhead{Reference} }
\startdata
Milky Way warp	& 1.7--$8.6\times 10^{12}$	& $l=270\degr$, $b=0\degr$
	& \nodata\\
Sgr dSph	& 1.6--$7.3\times 10^{12}$	& $l=276\degr$, $b=0\degr$
	& 1, 2\\
LMC		& $2\times 10^{14}$		& $l=184\degr$, $b=9\degr$
	& 3\\
Fornax dSph	& $3\times 10^{12}$		& $l=106\degr$, $b=-16\degr$
	& 4, 5\\
Ursa Minor dSph	& $3\times 10^{11}$		& $l=213\degr$, $b=9\degr$
	& 4, 6\\
Sculptor dSph	& $1\times 10^{11}$		& $l=226\degr$, $b=7\degr$
	& 4, 7\\
\enddata
\tablerefs{(1) Ibata et al. 1997; (2) Helmi \& White 2001;
(3) Kroupa \& Bastian 1997; (4) Mateo 1998;
(5) Piatek et al. 2002; (6) Schweitzer, Cudworth, \& Majewski 1997;
(7) Schweitzer et al. 1995}
\end{deluxetable}
Table~\ref{satellites table} shows the magnitude and direction of the
angular momenta of the Galactic satellites with measured proper
motions, along with that of the Milky Way warp which is calculated
in Section~\ref{Lwarp section}.
The orbital angular momentum of the Large Magellanic Cloud (LMC)
was calculated using data from \citet{kroupa and bastian97}.
For the remaining satellites, the mass was taken from \citet{mateo98} and
the velocity vector from the tabulated reference.

\section{The angular momentum associated with the Milky Way warp}
\label{Lwarp section}
\begin{deluxetable}{lcccc}
\tablecaption{Disk parameters\label{disk table}}
\tablehead{ & \multicolumn{4}{c}{Model} \\ & \colhead{1} & \colhead{2} & \colhead{3} & \colhead{4}}
\startdata
\sidehead{\citet{dehnen and binney98}:}\\
Stellar disk scale length $R_{d,*}$ (kpc)	
	& 2.0 & 2.4 & 2.8 & 3.2 \\
ISM disk scale length $R_{d,\mathrm{ISM}}$ (kpc)	
	& 4.0 & 4.8 & 5.6 & 6.4 \\
Surface density at solar circle $\Sigma_0$ ($M_{\Sun}~\mathrm{pc^{-2}}$)
	& 43.3 & 52.1 & 52.7 & 50.7 \\
\sidehead{Derived warp angular momenta: ($10^{12}~M_{\Sun}~\mathrm{kpc~km~s^{-1}}$)}\\
Stellar disk
	& 1.10 & 2.16 & 3.36 & 4.73 \\
ISM disk
	& 0.52 & 1.10 & 1.78 & 2.57 \\
Total
	& 1.62 & 3.26 & 5.14 & 7.30 \\
Direction
	& \multicolumn{4}{c}{$l=270\degr\pm 10\degr$, $b=0\degr$}\\
\enddata
\end{deluxetable}
I calculate the component of the disk angular momentum which
is due to the warp in the Milky Way's disk,
i.e., that which is not directed toward
the North Galactic Pole (NGP). If the disk rises
a height $h(R)$ above the plane at cylindrical radius
$R$, then the total angular momentum in the disk which
is due to the warp is
\begin{equation}
\label{Lw integral}
L_w = \int_{R_w}^\infty 2 \pi R^2 v_c \Sigma(R) 
	\frac{h(R)}{\sqrt{h(R)^2 + R^2}} \, dR.
\end{equation}

The mass distribution of the
disk is taken from \citet{dehnen and binney98}. The disk
surface density for a given component in these models is given by
\begin{equation}
\Sigma(R) = \Sigma_d \exp\left( -\frac{R_m}{R} -\frac{R}{R_d}\right),
\end{equation}
where $\Sigma_d$ is the normalization, $R_d$ is the scale length
of the component, and $R_m$ is introduced to allow the ISM to have
a central depression\footnote{Note that 
equation~(1) of \citet{dehnen and binney98} has
a typo which is fixed above
(W.~Dehnen 2002, private communication)}.
$R_m=4$~kpc for the gas disk and $R_m=0$ for the stellar disk.
The relative contributions to the surface density at the solar
circle $\Sigma_0$ are 0.25 for the ISM and 0.75 for the stars.
\citet{dehnen and binney98} distinguish between thin and thick
disk components of the stellar disk, but because these only
differ in vertical scale height, which does not affect the angular
momentum, I treat them as a single component.
Their models 1--4, which differ primarily in disk scale length, $R_d$,
are all acceptable fits to the observations, and therefore
provide a reasonable range of mass distributions with which to
estimate the angular momentum. Table~\ref{disk table}
gives the essential parameters for the four models.

The circular velocity, $v_c$, of the disk from 3~kpc to the solar
circle is $\approx 200~\mathrm{km~s^{-1}}$ \citep[e.g.,][]{merrifield92}.
While most measurements at $R>R_0$ show a rising rotation curve,
\citet{binney and dehnen97} argue that a constant rotation curve
is consistent with the data when the correlations between errors
are taken into account. I adopt $v_c=200~\mathrm{km~s^{-1}}$ at all
radii. The uncertainty in the angular momentum due to uncertainties
in the mass models dominates over any error in the circular velocity.

The height of the warp above the plane as a function of radius, $h(R)$, appears
to differ for the stars and for the gas. \citet{drimmal et al00}
fit Hipparcos measurements of OB stars and find
\begin{equation}
h(R) = \cases{ (R-R_w)^2/R_h & $R>R_w$\cr 0 & $R\le R_w$\cr},
\label{drimmal h(R)}
\end{equation}
with the warp starting at $R_w=6.5$~kpc and scaled by $R_h=15$~kpc.
\citet{BM} approximate the $m=1$ mode of the
ISM warp as
\begin{equation}
h(R) = \cases{ (R-R_w)/a & $R>R_w$\cr 0 & $R\le R_w$\cr},
\label{BM h(R)}
\end{equation}
where $R_w=10.4$~kpc and $a=5.6$ when converted to $R_0=8$~kpc as assumed
in the \citet{dehnen and binney98} models \citep{tsuchiya02}.
\citet{BM} also fit an $m=2$ mode, but the net
angular momentum of any even $m$ mode is aligned with the angular momentum
of the flat disk, so it will not contribute.

I use equation~(\ref{drimmal h(R)}) for the stellar disk and
equation~(\ref{BM h(R)}) for the gas disk. The results are shown in
Table~\ref{disk table}.
The majority of the angular momentum is contained in the range
$10\la R\la 25$~kpc in all models. The Sun lies within
$10\degr$ of the line of nodes, so $L_w$ is directed toward
$260\degr\la l\la 280\degr$, $b=0\degr$.

\section{Discussion}
\label{discussion section}
Table~\ref{satellites table} shows the magnitude and direction
of the angular momentum of the Milky Way warp and of the Galactic
satellites with measured proper motions.
Both the magnitude and direction of the angular momentum of Sagittarius
are strikingly similar to
that of the Milky Way warp. There
is no \textit{a priori} reason to expect this; the angular momenta of the other
satellites with known orbits span three orders of magnitude
and almost $180\degr$ of galactic longitude (although there is a strong
tendency for the satellites to have polar orbits with low values of
$|b|$, as suggested by \citet{lynden-bell76} and noted in the
anisotropic distribution about the Milky Way by \citet{hartwick00}).
The coincidence of the two angular momentum vectors
is probable evidence that they are dynamically coupled, i.e., that
Sagittarius is the perturber responsible for the Galactic warp.

There are three possibilities for the nature of the coupling.
The first is a direct gravitational tidal torque
by the satellite itself \citep{hunter and toomre69}, the second
is the gravitational torque of a wake in the Galactic dark matter
halo \citep{weinberg98,tsuchiya02}, and the third is an impulsive
deposition of momentum to the gas disk by passage through it
\citep{ibata and razoumov98}. The direct tidal torque for a satellite
of mass $m$ and distance $r$ scales as $m/r^3$. Therefore, the
direct tidal effect of Sgr is no stronger than that of the LMC,
whose direct tidal torque is not sufficient to induce the warp
\citep{hunter and toomre69}.
This means that the gravitational torque of Sgr itself cannot
be the coupling mechanism.

If the primary perturber is instead a wake in the halo, the strength
of the torque scales as $m_{\mathrm{wake}}/r_{\mathrm{wake}}^3$.
The mass of the wake scales as the mass of the satellite and as the
density of the halo at the wake radius \citep{weinberg98}.
The wake develops at half the satellite's orbital radius \citep{tsuchiya02}.
Therefore, for an isothermal halo, the strength of the torque scales
as $m/r^5$. In this case, the effect of Sagittarius is 10--50 times
stronger than that of the LMC. It is plausible that in
\citet{tsuchiya02}'s lower mass simulation, in which the LMC did
not excite a warp, a satellite with Sagittarius's parameters
would have. Further simulations which better reproduce the observed
Sgr-MW system could
confirm or falsify this suggestion.

\citet{ibata and razoumov98} suggest that the impulsive deposition
of momentum to the gas disk could excite the warp. The mass they
use for Sgr, $5\times 10^9~M_{\Sun}$, is quite large,
and they find very little warping in their $1\times 10^9~M_{\Sun}$
simulation. However, they only model a single interaction. In order
for the angular momenta to reach an equilibrium, as they appear
to have done, there must be repeated or continual encounters.
\citet{helmi and white01} find orbital periods of $\sim1$~Gyr for
Sagittarius, indicating that it has passed through the disk
several times.
Further simulations that follow the evolution of
the system over many encounters are necessary to better understand
the predictions of this model; meanwhile, it cannot be ruled out.

\section{Summary}
The orbital angular momentum of the Sagittarius dwarf galaxy and
the component of the Milky Way disk angular momentum due to the
Galactic warp are both directed toward $l\approx 270\degr$,
$b=0\degr$ with magnitude 2--$8\times 10^{12}~M_{\Sun}~\mathrm{kpc~km~s^{-1}}$.
Such a coincidence suggests that they are a coupled system,
i.e., that Sgr is responsible for the warp. The direct gravitational
tidal torque of Sgr cannot cause the warp. Interaction via a gravitational
wake in the Milky Way's dark matter halo, and impulsive deposition of
momentum into the disk by passing through it
are still both possible coupling mechanisms.
More simulations of each of these models are necessary to discriminate
between their effects.

\acknowledgements{Many thanks to Casey Meakin for useful
discussions and comments.}

\end{document}